\documentclass[aip,graphicx]{revtex4-1}

\usepackage{graphicx}
\usepackage{subfigure}
\usepackage{amssymb}
\usepackage{amsmath}
\usepackage{gensymb}
\usepackage{hyphenat}

\draft 

\begin{document}


\title{Magnetic superlens-enhanced inductive coupling for wireless power transfer} 



\author{Da Huang}
\affiliation{Center for Metamaterials and Integrated Plasmonics, Duke University, Durham, North Carolina 27708, USA}

\author{Yaroslav Urzhumov}
\affiliation{Center for Metamaterials and Integrated Plasmonics, Duke University, Durham, North Carolina 27708, USA}

\author{David R. Smith}
\email{drsmith@ee.duke.edu}
\affiliation{Center for Metamaterials and Integrated Plasmonics, Duke University, Durham, North Carolina 27708, USA}

\author{Koon Hoo Teo}
\affiliation{Mitsubishi Electric Research Laboratories, 201 Broadway St. $8^{th}$ floor, Cambridge, Massachusetts 02139, USA}

\author{Jinyun Zhang}
\affiliation{Mitsubishi Electric Research Laboratories, 201 Broadway St. $8^{th}$ floor, Cambridge, Massachusetts 02139, USA}


\date{\today}

\begin{abstract}
We investigate numerically the use of a negative-permeability "perfect lens" for enhancing wireless power transfer between two current carrying coils. The negative permeability slab serves to focus the flux generated in the source coil to the receiver coil, thereby increasing the mutual inductive coupling between the coils. The numerical model is compared with an analytical theory that treats the coils as point dipoles separated by an infinite planar layer of magnetic material [Urzhumov \textit{et al.}, Phys. Rev. B, \textbf{19}, 8312 (2011)]. In the limit of vanishingly small radius of the coils, and large width of the metamaterial slab, the numerical simulations are in excellent agreement with the analytical model. Both the idealized analytical and realistic numerical models predict similar trends with respect to metamaterial loss and anisotropy. Applying the numerical models, we further analyze the impact of finite coil size and finite width of the slab. We find that, even for these less idealized geometries, the presence of the magnetic slab greatly enhances the coupling between the two coils, including cases where significant loss is present in the slab. We therefore conclude that the integration of a metamaterial slab into a wireless power transfer system holds promise for increasing the overall system performance.
\end{abstract}


\maketitle 


\section{Introduction}
Since the earliest suggestions of wireless power transfer (WPT) by Tesla in 1891\cite{NOLIT1978}, the concept of conveying electromagnetic energy from one location to another without wires has captured enormous interest \cite{Matsumoto2004,Mankins2011,Deyle2008}. Stimulated by the growing demand in mobile electronic devices and electric vehicles, numerous efforts are now underway to develop new configurations capable of transporting energy wirelessly, and generally seeking to increase the efficiency of all WPT schemes as much as possible \cite{Cannon2009,Chinga2009,Wang2011,Reinhold2007,Yu2011}. WPT techniques have crossed over from research to product development, with several commercial units being offered for charging low-power consumer \cite{W.1970,Zink2002,Bruning2008} and medical electronics \cite{Ramrakhyani2011}. The majority of contemporary WPT schemes employ a near-field based, inductive coupling mechanism to achieve high transfer efficiency over short distances ($1-5$ cm); for acceptable efficiency in these schemes, the distance between the source and receiver must be considerably smaller than the dimension of the transmitter or receiver\cite{Waffenschmidt}. This proximity requirement makes the integration of WPT protocols into many electronic systems difficult or unfeasible, since the transfer efficiency in the near-field schemes decreases as a high-exponent power law with the transfer distance. The transfer efficiency is defined here as the fraction of power delivered to the resistive load relative to the total consumed power.

Recently, a WPT system with a relatively high transfer efficiency over a moderate propagation range has been developed, based on a pair of inductively coupled, resonant coils with large quality factors (Q-factors)\cite{Kurs2007}. In subsequent experiments performed at Intel, the efficiency in this resonant WPT system was shown to be as high as $75\%$ over a distance of $0.5$m between the source and receiver coils \cite{Sample2011}. Two key factors in the resonant WPT system determine the transfer efficiency: (1) the Q-factors of the resonators and (2) the mutual coupling strength. Strong coupling between the two resonators increases the energy exchange rate, thus increasing the power transfer efficiency. Though high-Q resonant WPT systems can have excellent power transfer efficiency, they are usually not preferred in practical applications since the coils tend to be more sensitive to the environment, and dynamic control is difficult to implement \cite{Si2008}. Nevertheless, given the demand for wireless powering and charging of devices, the efficiency associated with WPT between resonators motivates the continued exploration and improvement of such systems.

An obvious means of increasing the efficiency of an inductive, resonant WPT system is to increase the mutual inductive coupling between the source and the receive resonators. Because the distance between the source and receiver is so much smaller than the wavelength, the relevant field distribution is quasistatic and the inductive coupling relates predominantly to the amount of magnetic flux emanating from one coil that is captured by the second coil. Enhancing coupling efficiency equates to modifying this field distribution, which in turn means focusing or otherwise controlling the near fields.

A convenient means of controlling fields is through the use of a material. High permeability magnetic \nohyphens{materials}, for example, can guide magnetic flux and are used in motors and transformers \cite{Hasegawa2004,Wiltshire2001}. Inherently magnetic materials (i.e., those that derive their magnetic properties from electron spin), however, tend to be heavy and cumbersome, and have limited ranges of applicability. Additionally, one has very little control over their intrinsic loss tangents. Over the past decade, the concept of artificial magnetism--using inductive elements to mimic \nohyphens{magnetic} media--has been introduced and achieved widespread interest \cite{Wiltshire2001,Freire2008c,Shamonina2004}. In particular, the use of artificially structured metamaterials with negative effective permeability and permittivity has led to new opportunities for managing near-fields, as exemplified by the "perfect" lens. The perfect lens, introduced by Veselago in 1968\cite{Veselago1968} and further developed by Pendry \textit{et al.} in 2000\cite{Pendry2000}, is a planar slab whose electric permittivity and magnetic permeability both assume the values of -1 at a given frequency; a source placed on one side of the perfect lens is reproduced  as an image on the other side of the slab, with both far- and near-fields refocused to the image\cite{Pendry2000,Cummer2003,Merlin2007,Grbic2008,Smith2003a,Fang2005a,Cui2005}.

In the WPT context, the perfect lens geometry is a good starting point for improving the efficiency of coupling between two coils. Viewing the first coil as a source, any loss in transfer efficiency to the receiver can be explained using the concept of magnetic flux divergence, which is bound to happen at distances greatly exceeding the dimensions of the source. A perfect lens which captures the near-fields should be able to refocus the flux at the receiver, thus enhancing the coupling. The use of a perfect lens to improve WPT efficiency was considered by Mitsubishi Electric Research Lab \cite{Teo2011}, who, in recent experiments \cite{Wang2011}, demonstrated efficiency gains in a resonant WPT system. A rigorous analysis of the coupling enhancement offered by a perfect lens situated between two magnetic dipoles was subsequently performed by Urzhumov \textit{et al.} \cite{Urzhumov2011}. One of the key conclusions of that work was that the efficiency increase through enhanced mutual coupling could prevail over the reduction in efficiency caused by material losses, assuming realistically large resistive loads on the receiver side. This conclusion is critical for the consideration of metamaterials in WPT schemes, since metamaterials are frequently formed using conducting circuits that can exhibit significant Ohmic losses.

The analysis published by Urzhumov \textit{et al.} \cite{Urzhumov2011} employed a simplified geometry--such as an infinitely large slab and point dipoles for source and receiver--to obtain closed-form, analytical expressions that would provide initial insights into the possibility of efficiency gains with metamaterials. Here, we make use of full-wave, finite-element based simulations to analyze the mutual coupling between finite-diameter coils situated on either side of a finite-sized planar lens. We also make use of the above-mentioned prior analytical results to benchmark our numerical simulations in the limiting case of vanishingly small coils and very large-width slabs. After confirming the agreement between simulation and theory for several test cases, we next consider more realistic geometries, investigating the effects of finite coil diameter, finite slab width, and material loss and anisotropy.

\section{The Numerical Model}
We analyze the mutual inductance between the coils numerically by evaluating the magnetic flux in the receiver coil generated from the source coil in COMSOL Multiphysics. Here we model the resonant coil by a finite-size coil composed of a single turn of conducting wire, which forms a magnetic dipole. The magnetic dipole approximation is applicable to this geometry \nohyphens{since} the modes corresponding to deeply sub-wavelength, high-Q coil resonators, such as spirals or solenoids, are predominantly magnetic-dipolar in nature.

In general terms, the efficiency $\eta$ of a WPT system consisting of a source and a receiver coil is \cite{Urzhumov2011}

\begin{equation}\label{eqn:eta}
  \eta=\frac{P_2^0}{P_1+P_2}=\frac{R_2}{R_2^{\rm{eff}}}\frac{\chi}{1+\chi}
\end{equation}

where

\begin{equation}\label{eqn:chi}
  \chi=\frac{P_2}{P_1}=\frac{R_2^{\rm{eff}}\omega^2|L_{\rm{21}}|^2}{R_1^{\rm{eff}}|Z_2|^2}.
\end{equation}

\noindent Here $P_2^0$is the power extracted from the WPT system at the receiver end; $P_1$ and $P_2$ represent the power dissipated at the source and receiver coil, in which the power dissipated by the load presents as loss, and both the Ohmic loss in the system and the radiation loss from coils are included; $R_2$is the resistive load at the receiver coil; $R_2^{\rm{eff}}$  and $R_1^{\rm{eff}}$  are the effective resistances at the receiver and source coils, respectively (both include the loss from the self and mutual inductances); $\omega$ is the operating frequency of the system; and $Z_2$ is the effective impedance of the receiver coil, including the self-inductance and capacitance. After having optimized all other factors, Eqs.\ref{eqn:eta} and \ref{eqn:chi} show that the transfer efficiency increases quadratically with $L_{\rm{21}}$.

\begin{figure}
  \includegraphics[scale=1]{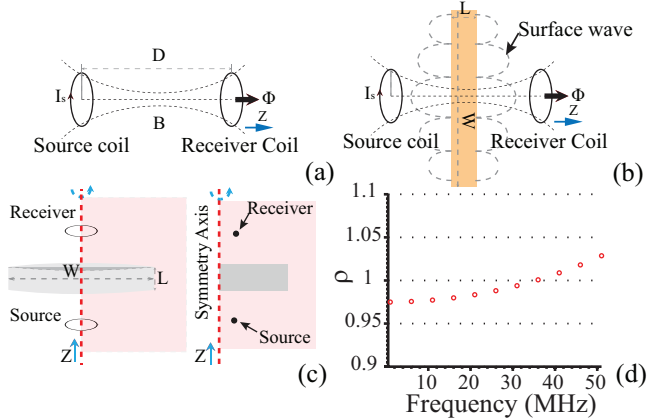}\\
  \caption{(color online)(a) Two-coil coupling system; (b) Two-coil coupling system with MM lens; (c) The 3D system configuration with MM slab and its equivalent axisymmetric 2D model in COMSOL; (d) The ratio between the numerically calculated and analytically predicted mutual inductance in the configuration shown in Fig.\ref{fig:1}(a).}\label{fig:1}
\end{figure}

The mutual coupling between the two coils can be evaluated by numerically calculating the mutual inductance between them, which is the ratio of the induced voltage on the receiver coil and the excitation current on the source coil, or

\begin{equation}\label{eqn:L21}
  L_{21}=\frac{V_{\rm{induced}}}{I_s}=\frac{-j\omega\int_{S_2}\vec{B}\vec{n} dA}{I_s}=\frac{-j\omega\oint_{\partial S_2}\vec{E} d\vec{l}}{I_s}
\end{equation}

\noindent where $\vec{B}$ is the magnetic flux density which can be computed in full wave simulations; $\vec{n}$ is the unit vector normal to the surface of $S_2$ ; $\vec{E}$ is the electric field; $\vec{l}$ is tangent vector to the loop of infinitesimally thin wire, and $S_2$ is the area encircled by the receiver coil. The two-coil system diagram is shown in Fig.\ref{fig:1}(a).

The presence of a metamaterial lens is simulated by including a slab of material with finite diameter $W$ and thickness $L$ placed between the two coils, as in Fig.\ref{fig:1}(b). The dominant field components are magnetic, since $D\ll\lambda_0$ , where $\lambda_0$ is the free space wavelength and hence the system operates in quasi-magnetostatic regime.. For this reason, it is expected that only the magnetic response of the slab will impact the near-field focusing, and thus only a magnetic component is assumed in this study. The magnetic slab is also desirable since it is easier to fabricate a metamaterial where fewer elements of the constitutive tensors need to be controlled. We consider then magnetic \nohyphens{slabs} with diagonal material properties, where the material properties can be represented as $\mu_{\rm{eff}}=[\mu_{\rm{z}},\mu_{\rm{y}},\mu_{\rm{z}}]$,and $\epsilon_{\rm{eff}}$  is irrelevant. For simplicity, we assume $\epsilon_{\rm{eff}}=1$.

To perform the preliminary comparisons between the numerical and analytical models, an axisymmetric two-dimensional geometry is simulated, in which the current coil is represented by an out of plane line current, shown in Fig.\ref{fig:1}(c). As shown, the axisymmetric geometry assumes a finite-diameter, cylindrical disk (which we refer to as a "slab" or "lens" in this paper). Note that this rotational symmetry only exists for the $z$-oriented dipole shown in Fig.\ref{fig:1}(a), which is the second case studied in Ref.\onlinecite{Urzhumov2011}. For the other dipole orientations, such as $x$-oriented dipole, this rotational symmetry is not present and the presented numerical model is no longer valid (a full 3D model then can be built for arbitrary dipole orientation analysis\cite{Ghali2009}). However, the system shown in Fig.\ref{fig:1}(a) is an optimal configuration since the coupling is a factor of two stronger than for the first dipole orientation case considered in Ref. \onlinecite{Urzhumov2011}. The magnetic flux density B can be precisely determined over the surface enclosed by the receiver coil through a full wave simulation.  Thus, we can numerically evaluate the mutual inductance between the two coils since all the terms in Eq.\ref{eqn:L21} can be determined in the simulator.

In Ref. \onlinecite{Urzhumov2011}, the mutual coupling between two magnetic point dipoles is studied analytically in the presence of a negative permeability slab. The mutual inductance is calculated and presented as a closed form expression. The resonant coil is first approximated as a single coil, finite-sized coil with a constant current and infinitesimal coil cross section area, and then further reduced to a magnetic point dipole with equivalent magnetic dipole moment $M=I_s A=\pi R^2 I_s$ . In the present numerical study, the resonant coils are also approximated by finite size single coil, but are not further reduced to point dipoles. The analytical expression for the mutual inductance in the presence of the slab is:

\begin{equation}\label{eqn:L21Ana}
  L_{21}=-\frac{\mu_0 \pi R^4}{2}\frac{4\alpha/\mu_x}{a(2\alpha L)^3}\Phi_L\left(-b/a,3,\frac{\alpha L+D-L}{2\alpha L}\right)
\end{equation}

\noindent where $\mu_0$ is the free space permeability; $\mu_x,\mu_y,\mu_z$ is the effective permeability of the MM lens; $\alpha=\sqrt{\mu_x/\mu_z}$ ; $a=-(\alpha/\mu_x+1)^2$ ; $b=(\alpha/{\mu_x}-1)^2$ ; and $\Phi_L$ is the standard Lerch transcendent function, as defined in Wolfram Mathematica software and Gradshteyn-Ryzhik \cite{Grover1982}.

In the absence of the slab when $[\mu_x,\mu_y,\mu_z]=[1,1,1]$, the mutual inductance between the two coils reduces to $L_{21}^{\rm{vac}}=-\mu_0 \pi R^4/2D^3$ , when the two coils have the same radius $R$. In the following, we will use the enhancement factor defined as the ratio between the calculated mutual inductance and theoretically predicted mutual inductance in vacuum as the criterion by which to evaluate the enhancement of mutual coupling that occurs in the presence of the slab.

\begin{equation}
    \rho=\frac{L_{21}}{L_{21}^{\rm{vac}}}
\end{equation}

We first investigate numerically the retardation effect associated with the mutual inductance between the \nohyphens{two} coils in free space. The analyzed system consists of \nohyphens{two} coils, each with radius $R=0.01$ m and separated by a distance  $D=0.5$ m. the frequency of the source is swept from $5$ MHz to $50$ MHz, corresponding to wavelengths between $6$ m to $60$ m. At the smallest wavelength in the frequency bandwidth, the radius of the coil is less than $0.2\%$ of wavelength, and the transfer distance is less than $10\%$ of the wavelength; we thus identify this configuration as a sub-wavelength WPT system. As shown in Fig.\ref{fig:1}(d), the mutual coupling calculated from the numerical model is essentially the same as predicted by the analytical model, deviating by less than $2\%$ of over the broad frequency range. Any discrepancy between the numerical and analytical models is partially attributable to the finite radius of the wires, finite diameter of the coils, and also the numerical error from the applied FEM simulator.

In a practical WPT configuration, the source and receiver(s) are solenoids or spirals of finite diameter. Also, any metamaterial structure that implements the negative permeability lens will have finite extent in the transverse direction. In the following section, we investigate the impact of the finite size of the coils and the slab. If not otherwise specified, the frequency of operation is $10$MHz, consistent with the frequency band used in previous experimental investigations and some commercial implementations \cite{Wang2011,Kurs2007,Sample2011}; the radius of the source and receiver coils is $R=0.01$m; the transfer distance is $D=0.5$m; the diameter of the negative permeability slab is $W=2$m and its thickness $L=D/2=0.25$m. Note the thickness corresponds to the "perfect lens" condition \cite{Pendry2000}.

\section{Finite size and loss effects in isotropic-permeability lens}

\begin{figure}[h,b]
\centering
  \includegraphics[scale=1]{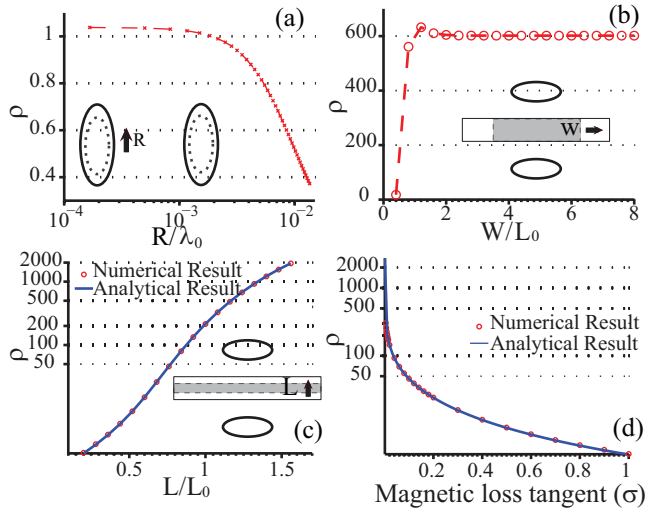}\\
  \caption{(color online)(a) The ratio of simulated $L_{21}$ to the theoretical  $L_{21}$ calculated in the point-dipole limit, as a function of coil radius $R$ normalized to the wavelength (coil retardation parameter); (b) mutual inductance enhancement as a function of the MM lens width $W$; (c) mutual inductance enhancement versus lens thickness for a fixed inter-coil distance, where $L_0=D/2$ is the optimum lens thickness; (d) magnetic loss effect on mutual inductance enhancement. }\label{fig:2}
\end{figure}

We first investigate the effect of finite coil size on the mutual inductance in the absence of the slab. In the analytical study, coils were approximated as point dipoles, while in the numerical model the coils have finite radius $R$. To study the impact of finite radius coils, we vary the radius in the simulation, starting from a very small radius (relative to the wavelength). As a function of increasing radius, the computed mutual inductance deviates from the analytical model, as expected. As shown in Fig.\ref{fig:2}(a), when the coil size reaches roughly $0.001\lambda_0$, where $\lambda_0$ is the free space wavelength corresponding to the source frequency ($10$ MHz), the dipole approximation used in the analytical calculations is no longer accurate.

The comparison in Fig.\ref{fig:2}(a) shows that the deviation in the mutual inductance between two finite size coils and two idealized dipole remains within $15\%$ when the coil radius is approximately $0.4\%$ of the free space wavelength, or a quarter of the transfer distance.  Further increasing the coil radius decreases the mutual inductance to the point that Eq.\ref{eqn:L21} no longer provides an accurate estimate for the mutual inductance. For a $10$ MHz drive frequency, the free space wavelength is $30$m; thus, Eq.\ref{eqn:L21} is valid when the coil radius is smaller than $12$cm. This condition is maintained throughout the following analysis, where a coil radius of $1$ cm is modeled unless otherwise specified.

A practical question that arises relates to the necessary size of the negative permeability slab. While the analytical model assumed a slab infinite in the transverse directions, a physical implementation would need to be finite, and preferably as small as possible. The "perfect lens" material condition in the magnetostatic limit, which presumably leads to a perfect refocusing of the magnetic near-fields, is $[\mu_x,\mu_y,\mu_z]=[-1,-1,-1]$\cite{Pendry2000}.  For ideal condition, the lens has a thickness $L=D/2$, and the two coils have equal distance $D/4$ to either surface of the lens, being placed coaxially at the object and image planes. Note that the perfect lens condition was originally derived without consideration to there being any power transfer between source and receiver \cite{Pendry2000}; however, even when strong coupling persists between the source and receiver, the same conditions remain approximately optimal \cite{Urzhumov2011}.

To examine the effect of a finite-sized perfect lens on the mutual inductance, we simulate a finite width slab with the real part of the permeability tensor assigned the values $[\mu_x,\mu_y,\mu_z]=[-1,-1,-1]$. A small imaginary part $0.001j$  is added to each of the three diagonal components of the permeability tensor to facilitate numerical convergence of the finite-element scheme with the mesh size, since the exact perfect lens condition can be numerically unstable due to the presence of strong surface modes with arbitrarily small wavenumbers;  finite damping suppresses high-k surface modes, making it possible to achieve accurate FEM results. Thus, the permeability used in the simulations is isotropic and has the value $\mu=-1-0.001j$. Here, we use the time dependence convention $E \sim \exp(j\omega t)$, for compatibility with COMSOL Multiphysics. Strong enhancements in the mutual inductive coupling have been predicted in the analytical analysis, which is also revealed in the numerical results discussed below.

As shown in Fig.\ref{fig:2}(b), the enhancement factor $\rho$ is, as expected, a function of $W$. $\rho$ increases rapidly and saturates when $W \approx D$. In this sub-wavelength regime, the field excited along the MM lens surface is a magnetostatic surface resonance (MSR)\cite{Urzhumov}, which provides some insight as to the mechanism of the enhancement. The source coil excites a MSR characterized by near-fields that oscillate rapidly across the slab surface and build up strongly on the surface but decay exponentially away from it. The surface field pattern creates a focus of the near-fields at the optimal "focus" position \cite{Pendry2000,Cummer2003,Shvets2003,Wiltshire2001,Freire2010,Merlin2007}, much in the same way that fields in the aperture of a lens can produce a focus. The important difference between conventional lenses and the superlens, however, is that the latter confines the fields in the focus only transversely, forming a two-dimensional "focal line" aligned with the point source, along which the fields decay roughly exponentially. One would expect that, below a certain diameter, the superlens would lose its ability to produce a perfect, predictable two-dimensional focus; in particular, when the width of the lens is small, the field excited on the edge of the lens can produce uncontrolled focusing or defocusing. In terms of the surface resonance theory, a finite-diameter superlens has a spectrum of surface modes that is spread over a range of permeability values, in contrast with an infinite-diameter lens whose spectrum has an eigenvalue accumulation point at $\mu=-1$, and thus all MSR modes with wavenumbers $k\gg1/L$, where $L$ is the lens thickness, can be efficiently and simultaneously excited when $\mu=-1$. Exciting surface modes with a range of wavenumbers simultaneously is a key ingredient to perfect near-field imaging; a slab whose width $W$ is not much larger than its thickness $L$ does not support a quasi-continuous range of resonant transverse wavenumbers for any given negative value of $\mu$.

The effects of the finite size of the slab can be seen in Fig.\ref{fig:2}(b), where the enhancement factor becomes larger than its asymptotic value for certain slab widths. The enhancement factor saturates for larger widths, where any influence of the slab termination on the surface resonance modes becomes negligible.

While the perfect lens condition places a strict limit on the thickness of the slab, we are not necessarily interested in creating a perfect image for the WPT application; thus, for WPT purposes, we might expect the mutual inductance to increase as the slab thickness is increased. Fig.\ref{fig:2}(c) shows that the field intensity in the receiver coil indeed increases as the thickness of the slab increases with the overall distance between the source and receiver coils fixed at $D=0.5$ m. The slab material properties remain the same in this study: $\mu=-1-0.001j,\epsilon=1$ . Not surprisingly, these calculations show that the closer the coils can be placed to the slab-where the local magnetic fields are very large-the greater will be their coupling. The distance dependence is rather sharp, since the surface modes from the slab decay exponentially away from the slab. From the other point of view, the effective distance between the two coils is $D-2L$ in the presence of isotropic MM lens, comparing to $D$ in the absence of the MM lens. Therefore mutual inductance is much larger when the MM lens is thicker since the mutual inductance is proportional to the inverse of distance cubed as discussed before. The results are shown in Fig.\ref{fig:2}(c).

In the expected realizations of the negative-$\mu$ lens, metamaterials that exhibit artificial magnetism will likely be employed, as they have been used in recent WPT experiments\cite{Wang2011}. Because these metamaterials are based on inductive elements, such as split ring resonators, they have generally significant loss tangents related to the Ohmic losses in the constituent current coils. The issue of loss is of particular concern for WPT applications, since the losses represent reduced overall transfer efficiency. To investigate the impact of losses in the slab, we assume an isotropic, lossy medium of the form $\mu_{\rm{eff}} =-1(1+j\sigma)$ and $\epsilon=1$, where $\sigma$ is the magnetic loss tangent of the slab which is the ratio between the imaginary and real part of the effective permeability. The radius for both the source and receiver coil is $0.01$ m; and the slab width is $2$ m; both dimensions chosen such that all finite size effects discussed above are negligible. The thickness of the slab correspond to the perfect lens condition: half of the separation distance, or $L=D/2=0.25$ m . Results from analytical theory and numerical model are shown in Fig.\ref{fig:2}(d).

Excellent agreement is again found between the analytical and numerical models, which both show that even with a loss tangent on the order of unity, the mutual inductance is still doubled. The enhancement factor $\rho$ has a magnitude that varies inversely with respect to that of the magnetic loss tangent $\sigma$.  In practice, a negative permeability metamaterials can reach a loss tangent on the order of $0.1$ or less when $\mu \approx -1$\cite{Chen2011}. In this case, the mutual inductance enhancement ratio $\rho$ can reach values in the range $10$ to $100$, as shown in Fig.\ref{fig:2}(d).

\section{Anisotropic, negative-definite permeability lenses}

For a slab made of isotropic negative permeability, the ideal thickness corresponds to $D/2$ , for which the source and receiver coil are placed at the object and image planes next to the slab surfaces. The slab thus occupies half of the space between the source and receiver, making it somewhat bulky and obtrusive. A means of reducing the thickness profile of the slab, without changing the field structure in any manner, can by determined by applying techniques from transformation optics\cite{Roberts2009}.In particular, a compression transformation along the optical axis (the direction normal to the slab or the $z$-direction in Fig.\ref{fig:1}(a,b,c)) results in a material whose permeability is constant everywhere, but is anisotropic.

\begin{figure}[h,t,f]
  \includegraphics[scale=1]{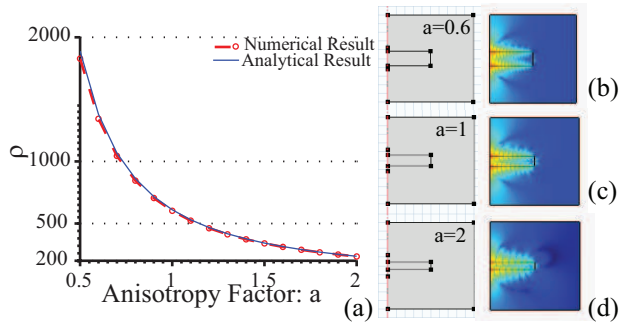}\\
  \caption{(color online)(a) Mutual coupling enhancement factor dependence on the anisotropy factor (same as lens compression factor); System configuration and magnetic field profile when (b) $a=0.6$, (c) $a=1$, (d) $a=2$.}\label{fig:3}
\end{figure}

As before, we consider a configuration of two coils separated by a distance $D=0.5$ m, with the slab placed in between the two coils. We introduce the anisotropy factor $a$  to indicate the spatial compression factor. Based on transformation optics theory, the permeability tensor required to implement a reduction in the slab profile from $L$ to $L/a$ is $[\mu_x,\mu_y,\mu_z ]=[-a,-a,-1/a]$ in order to reduce the lens thickness from $L_0$ to $L_0/a$\cite{Roberts2009}. We performed a parametric study, where $a$ sweeps from $0.6$ to $2$ and the lens thickness $L=D/(1+a)$ varies from $0.625D$ to $0.33D$. A constant imaginary part $\rm{Im}(\mu)=0.001$ has been added to all three components of the slab permeability. The parametric study result is shown in Fig.\ref{fig:3}(a).

Fig.\ref{fig:3}(a) shows that the enhancement factor is large when the anisotropy factor a is small, in which case the receiver coil is closer to the slab surface. When the receiver is closer to the slab, a stronger local field exists at the receiver coil due to the surface mode. Field plots of the slab surface mode when $a=0.6,1,2$ are shown in Fig.\ref{fig:3}(b, c, d). As the anisotropy factor $a$ increases, the lens thickness decreases, and both the receiver coil and source coil effectively move away from the slab surface. Even though the enhancement factor is reduced for larger values of $a$, the overall enhancement is still large, $\rho \approx 220$ at $a=2$, two orders of magnitude larger as compared with the coupling between two coils in the absence of the lens.

\begin{figure}[h,t,f]
  \includegraphics[scale=1]{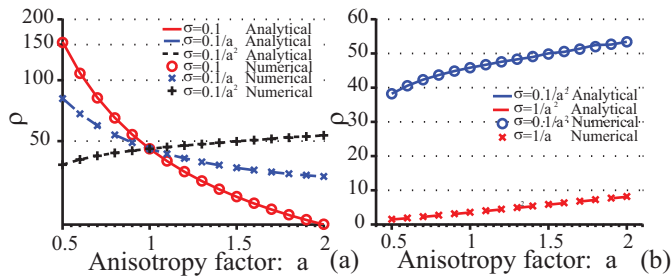}\\
  \caption{(color online) (a)Enhancement factor $\rho$ \nohyphens{depends} on the anisotropic MM lens's material loss tangent $\sigma=0.1, 0.1/a, 0.1/a^2$; (b) Mutual enhancement reduction while increase the material loss tangent from $\sigma=0.1/a^2$ to $\sigma=1/a^2$ }\label{fig:4}
\end{figure}

The change in the enhancement factor is initially surprising, since the coordinate transformation should \nohyphens{preserve} all properties of the field distribution perfectly. That is, the compressed slab with anisotropic medium should perform as well as the isotropic slab, with the enhancement factor $\rho$ remaining constant for any value of the anisotropy factor $a$. The basic transformation optics theory, however, is geometrical in nature and does not immediately apply to complex material parameters. We thus suspect that the reason for the variation in the mutual coupling enhancement relative to the transformation optics prediction lies in the loss tangent applied to the anisotropic medium. In order to preserve the same performance as isotropic lens, the loss in the anisotropic medium must be scaled appropriately. We investigated scaling the loss tangent of the slab in several different ways: $\sigma=0.1,0.1/a,0.1/a^2$ . Results for the different scaling are shown in Fig.\ref{fig:4}(a).  As shown, an anisotropic medium with $\sigma=0.1/a^2$ yields the expected flat ($a$-independent) enhancement factor $\rho$ .

Having identified the proper scaling for the loss tangent with respect to the anisotropy factor, we study the impact of losses on the anisotropic slab. In the following, all components of permeability tensor share the same loss tangent $\sigma$ . For $\sigma=1/a^2$ the enhancement factor $\rho$ reduces from a value around $50$ to a unit value. Both the analytical and numerical result is shown in Fig.\ref{fig:4}(b). In the practical case where $\sigma \sim 0.025$, the \nohyphens{enhancement} $\rho$ still achieves around $50$ even when the slab is only quarter of the transfer distance.

The results in the above study show that material loss needs to be well controlled in the anisotropic lens, whenever its thickness is less than the thickness of the canonical, isotropic lens (i.e. $D>L/2$ and $a>1$). In order to achieve the same magnitude of coupling with a lens with $a>1$, the magnetic loss tangent needs to decrease roughly as $1/a^2$. In other words, there is a trade-off between increasing the transfer distance of the lens relative to its thickness, and decreasing resistive loss in the metamaterial.

\section{Indefinite-permeability lenses}

In the previous section, we have analyzed the performance of anisotropic lens, which enables reduction of the MM lens's thickness in the power transfer direction. Now we study the system in which the MM lens is formed by an indefinite-permeability medium (or \emph{indefinite medium} for short), where one of the three principal values of the permeability tensor has different sign from the other two \cite{Schurig2005,Lee2007,Yao2011}. The motivation to apply indefinite medium is to simplify the MM lens configuration in reality. By eliminating one (or two) negative component in the permeability tensor, one (respectively, two) resonant unit can be removed from each array element in the periodic artificial composite\cite{Chen2011}. This reduces the fabrication complexity, and even allows one to reduce the space between magnetic resonators and stack them more densely for enhanced magnetic oscillator strength and better magnetic loss tangents in the negative-$\mu$ band \cite{Chen2011}.Equivalent performance of WPT system with anisotropic MM lens has been verified experimentally comparing to the WPT system with isotropic MM lens \cite{Wang2011}.

\begin{figure}[h,t,f]
  \includegraphics[scale=1]{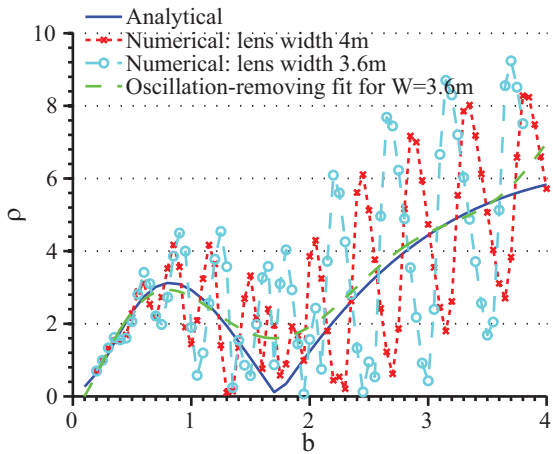}\\
  \caption{(color online)Indefinite-permeability lens: coupling enhancement as a function of $b=\rm{Re}({-\mu_{\rm{z}}})$.}\label{fig:5}
\end{figure}

An important distinction of the indefinite-medium lens is that it supports propagation of waves with wavenumbers greatly exceeding the free space wavenumber $k_0=\omega/c$, which is a consequence of the hyperbolic dispersion relation in an indefinite medium\cite{Smith2003,Alu2005,Schurig2003,Balmain2002,Jacob2006,Smith2004a,Smith2004b,Siddiqui2011,Cheng2008a}. Thus, even in the near field, spatial Fourier components that are evanescent in free space become propagating waves inside the indefinite medium. In our specific study, the anisotropic material properties for the MM lens is: $\mu_z=-b-0.1j, \,\mu_x=\mu_y=1$ and $\epsilon=1$. And the lens has a thickness $L=D/2$, with finite width in the numerical model, and infinite width in analytical model. The analytical result \nohyphens{predicts} the enhancement ratio is less than $10$ (the solid blue curve in Fig.\ref{fig:5}). Two indefinite MM lens with different lens width $W=3.6$m and $W=4$m are \nohyphens{studied} numerically. The numerical simulations, which utilize finite slab width $W$, reveal two superimposed oscillations in the $\rho$ vs $b$ curve.

The short oscillations in the numerical result are due to the Fabry-Perot resonances(FPR), which result from the finite width $W$ of the MM lens and are enabled by the above-mentioned hyperbolic dispersion property of the indefinite medium. Slab termination at a finite radius ($W/2$) creates additional reflecting planes and possibility for standing wave, which are not accounted for in the analytical model that assumes an infinitely wide slab.

To make a better comparison with the idealized analytical model, we have applied a curve fitting technique that eliminates these short oscillations. The fitting curve (dashed green curve in Fig.\ref{fig:5}) based on the numerical result matches well with the analytical prediction. In the fitted curve, as well as the analytical calculation, there is yet another, longer-period oscillation as a function of $b=-\rm{Re}(\mu_{\rm{z}})$; since those are captured by the idealized model, but are not present in the negative-definite lens case, we must attribute this longer-period oscillation to the FPR associated with the thickness $D$ of the lens. \nohyphens{Since} $D \ll W$ in all cases considered, FPR of the lens thickness $D$ are indeed expected to have a substantially longer period than FPR of the lens width W, which further confirms our explanation of both types of oscillations.

\begin{figure}[h,t]
  \includegraphics[scale=1]{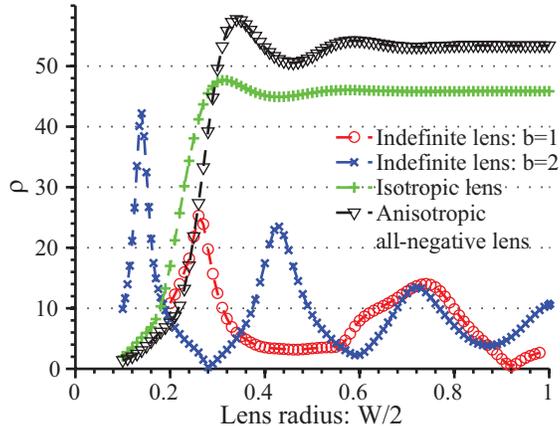}\\
  \caption{(color online)The summary of enhancement ratio obtained with isotropic MM lens, anisotropic MM lens and indefinite lens when the lens radius changes. Isotropic lens has $\mu=-1-0.1j$ and $L=D/2$ in which $D=0.5$m; Anisotropic lens has $[\mu_x,\mu_x,\mu_x ]=[-2-0.05j,-2-0.05j,-0.5-0.0125j]$ and $L=D/4$; Indefinite lens with $b=1$ has $[\mu_x,\mu_y,\mu_z]=[-1-0.1j,1,1]$ and $L=D/2$; Indefinite lens with $b=2$ has $[\mu_x,\mu_y,\mu_z ]=[-2-0.2j,1,1]$ and $L=D/2$. }\label{fig:6}
\end{figure}

Finally, Finally, we summarize typical enhancement ratios that can be obtained with isotropic, anisotropic all-negative lens (which is based on the transformation optics ideas), and two indefinite-medium lenses. The result is shown in Fig.\ref{fig:6}. In this summary, both isotropic and indefinite medium lens share the same loss tangent $\sigma=0.1$ and lens thickness $L=D/2=0.25$m. The anisotropic definite-negative permeability lens has loss tangent $\sigma=0.1/a^2$ and lens thickness $L=D/4=0.125$m since the anisotropy factor $a=2$.  In the indefinite medium lens, the oscillation in $\rho$ results from FPR modes associate with the finite dimension of the MM lens.

To summarize the results in Fig.\ref{fig:6}, we find that with a sufficiently small metamaterial loss, the inductive coupling enhancement provided by the negative-definite anisotropic permeability lens can be as substantial good as with an isotropic negative permeability lens of a substantially larger thickness. This result shows that anisotropic MM lenses can be potentially applied to reducing the size, volume and mass of the metamaterial component in the MM-enhanced wireless power systems.

With the indefinite permeability lens, optimum mutual coupling enhancement can be achieved by constructively rebuilding the field in the receiver coil and manipulating the MSR mode on the lens surface with the boundary effect. In particular, when the lens has width equal to $0.15$m, and the lens with $\mu_z=-2-0.2j$, the mutual coupling enhancement in the indefinite lens outperforms that of the isotropic and anisotropic MM lens when all three lenses share the same width.  The results shown here predict that, in some cases, the indefinite permeability lens can outperform isotropic or anisotropic negative-definite lenses when designed and optimized to take advantage of  the effects related to the finite slab size and boundary reflections.

\section{Conclusions}
We have analyzed the mutual inductive coupling enhancement by virtue of MM lenses added to the resonant near-field wireless power transfer systems. A few practical issues have been addressed here, including finite dimension of both the coil and MM lens, and the application of the anisotropic negative-definite and indefinite permeability lenses. In our results, the numerical result matches the analytical prediction with excellent precision whenever the numerical model operates within the assumptions of the analytical model. We find that the inductive coupling enhancement can be substantial even when the MM lens has significant resistive loss. The application of anisotropic, negative-definite permeability lenses requires fine control over the material loss, but enables the reduction of the metamaterial layer thickness. To make use of an indefinite permeability lens, the lens dimensions and permeability magnitude need to be carefully optimized in order to achieve the maximum possible enhancement; in certain regimes, the indefinite lens can perform as well as the definite-negative lens or even outperform the latter.

\begin{acknowledgments}

The authors are thankful to Matt Reynolds (Duke University) and Bingnan Wang (Mitsubishi Electric Research Laboratories) for useful discussions on wireless power transfer.
\end{acknowledgments}


\begin{thebibliography}{50}%
\makeatletter
\providecommand \@ifxundefined [1]{%
 \@ifx{#1\undefined}
}%
\providecommand \@ifnum [1]{%
 \ifnum #1\expandafter \@firstoftwo
 \else \expandafter \@secondoftwo
 \fi
}%
\providecommand \@ifx [1]{%
 \ifx #1\expandafter \@firstoftwo
 \else \expandafter \@secondoftwo
 \fi
}%
\providecommand \natexlab [1]{#1}%
\providecommand \enquote  [1]{``#1''}%
\providecommand \bibnamefont  [1]{#1}%
\providecommand \bibfnamefont [1]{#1}%
\providecommand \citenamefont [1]{#1}%
\providecommand \href@noop [0]{\@secondoftwo}%
\providecommand \href [0]{\begingroup \@sanitize@url \@href}%
\providecommand \@href[1]{\@@startlink{#1}\@@href}%
\providecommand \@@href[1]{\endgroup#1\@@endlink}%
\providecommand \@sanitize@url [0]{\catcode `\\12\catcode `\$12\catcode
  `\&12\catcode `\#12\catcode `\^12\catcode `\_12\catcode `\%12\relax}%
\providecommand \@@startlink[1]{}%
\providecommand \@@endlink[0]{}%
\providecommand \url  [0]{\begingroup\@sanitize@url \@url }%
\providecommand \@url [1]{\endgroup\@href {#1}{\urlprefix }}%
\providecommand \urlprefix  [0]{URL }%
\providecommand \Eprint [0]{\href }%
\providecommand \doibase [0]{http://dx.doi.org/}%
\providecommand \selectlanguage [0]{\@gobble}%
\providecommand \bibinfo  [0]{\@secondoftwo}%
\providecommand \bibfield  [0]{\@secondoftwo}%
\providecommand \translation [1]{[#1]}%
\providecommand \BibitemOpen [0]{}%
\providecommand \bibitemStop [0]{}%
\providecommand \bibitemNoStop [0]{.\EOS\space}%
\providecommand \EOS [0]{\spacefactor3000\relax}%
\providecommand \BibitemShut  [1]{\csname bibitem#1\endcsname}%
\let\auto@bib@innerbib\@empty
\bibitem [{\citenamefont {NOLIT}(1978)}]{NOLIT1978}%
  \BibitemOpen
  \bibfield  {author} {\bibinfo {author} {\bibnamefont {NOLIT}},\ }\href@noop
  {} {\emph {\bibinfo {title} {{Nikola Tesla Colorado Springs Notes
  1899-1900}}}}\ (\bibinfo  {publisher} {Nilola Tesla Museum},\ \bibinfo
  {address} {Beograd},\ \bibinfo {year} {1978})\ p.\ \bibinfo {pages}
  {395}\BibitemShut {NoStop}%
\bibitem [{\citenamefont {Matsumoto}\ \emph {et~al.}(2004)\citenamefont
  {Matsumoto}, \citenamefont {Hashimoto}, \citenamefont {Shinohara},\ and\
  \citenamefont {Mitani}}]{Matsumoto2004}%
  \BibitemOpen
  \bibfield  {author} {\bibinfo {author} {\bibfnamefont {H.}~\bibnamefont
  {Matsumoto}}, \bibinfo {author} {\bibfnamefont {K.}~\bibnamefont
  {Hashimoto}}, \bibinfo {author} {\bibfnamefont {N.}~\bibnamefont
  {Shinohara}}, \ and\ \bibinfo {author} {\bibfnamefont {T.}~\bibnamefont
  {Mitani}},\ }in\ \href@noop {} {\emph {\bibinfo {booktitle} {Proceedings of
  The 4th International Conference on Solar Power from Space}}}\ (\bibinfo
  {year} {2004})\BibitemShut {NoStop}%
\bibitem [{\citenamefont {Mankins}(2011)}]{Mankins2011}%
  \BibitemOpen
  \bibfield  {author} {\bibinfo {author} {\bibfnamefont {J.~C.}\ \bibnamefont
  {Mankins}},\ }\href
  {http://www.nss.org/news/releases/PressConference-IAA-SSP-Responses.pdf}
  {\enquote {\bibinfo {title} {{Space Solar Power The first International
  Assessment of Space Solar Power: Opportunities, issues and Potential Pathways
  forward}},}\ }\bibinfo {type} {Tech. Rep.}\ (\bibinfo  {institution}
  {International Academy of Astronautics},\ \bibinfo {year} {2011})\BibitemShut
  {NoStop}%
\bibitem [{\citenamefont {Deyle}\ and\ \citenamefont
  {Reynolds}(2008)}]{Deyle2008}%
  \BibitemOpen
  \bibfield  {author} {\bibinfo {author} {\bibfnamefont {T.}~\bibnamefont
  {Deyle}}\ and\ \bibinfo {author} {\bibfnamefont {M.}~\bibnamefont
  {Reynolds}},\ }\href {\doibase 10.1109/ROBOT.2008.4543341} {\bibfield
  {journal} {\bibinfo  {journal} {2008 IEEE International Conference on
  Robotics and Automation}\ ,\ \bibinfo {pages} {1036}} (\bibinfo {year}
  {2008})}\BibitemShut {NoStop}%
\bibitem [{\citenamefont {Cannon}\ \emph {et~al.}(2009)\citenamefont {Cannon},
  \citenamefont {Hoburg}, \citenamefont {Stancil},\ and\ \citenamefont
  {Goldstein}}]{Cannon2009}%
  \BibitemOpen
  \bibfield  {author} {\bibinfo {author} {\bibfnamefont {B.~L.}\ \bibnamefont
  {Cannon}}, \bibinfo {author} {\bibfnamefont {J.~F.}\ \bibnamefont {Hoburg}},
  \bibinfo {author} {\bibfnamefont {D.~D.}\ \bibnamefont {Stancil}}, \ and\
  \bibinfo {author} {\bibfnamefont {S.~C.}\ \bibnamefont {Goldstein}},\
  }\href@noop {} {\bibfield  {journal} {\bibinfo  {journal} {IEEE Transactions
  on Power Electronics}\ }\textbf {\bibinfo {volume} {24}},\ \bibinfo {pages}
  {1819} (\bibinfo {year} {2009})}\BibitemShut {NoStop}%
\bibitem [{\citenamefont {Low}\ \emph {et~al.}(2009)\citenamefont {Low},
  \citenamefont {Chinga}, \citenamefont {Tseng},\ and\ \citenamefont
  {Lin}}]{Chinga2009}%
  \BibitemOpen
  \bibfield  {author} {\bibinfo {author} {\bibfnamefont {Z.~N.}\ \bibnamefont
  {Low}}, \bibinfo {author} {\bibfnamefont {R.~A.}\ \bibnamefont {Chinga}},
  \bibinfo {author} {\bibfnamefont {R.}~\bibnamefont {Tseng}}, \ and\ \bibinfo
  {author} {\bibfnamefont {J.}~\bibnamefont {Lin}},\ }\href {\doibase
  10.1109/TIE.2008.2010110} {\bibfield  {journal} {\bibinfo  {journal} {IEEE
  Transactions on Industrial Electronics}\ }\textbf {\bibinfo {volume} {56}},\
  \bibinfo {pages} {1801} (\bibinfo {year} {2009})}\BibitemShut {NoStop}%
\bibitem [{\citenamefont {Wang}\ \emph {et~al.}(2011)\citenamefont {Wang},
  \citenamefont {Teo}, \citenamefont {Nishino}, \citenamefont {Yerazunis},
  \citenamefont {Barnwell},\ and\ \citenamefont {Zhang}}]{Wang2011}%
  \BibitemOpen
  \bibfield  {author} {\bibinfo {author} {\bibfnamefont {B.}~\bibnamefont
  {Wang}}, \bibinfo {author} {\bibfnamefont {K.~H.}\ \bibnamefont {Teo}},
  \bibinfo {author} {\bibfnamefont {T.}~\bibnamefont {Nishino}}, \bibinfo
  {author} {\bibfnamefont {W.}~\bibnamefont {Yerazunis}}, \bibinfo {author}
  {\bibfnamefont {J.}~\bibnamefont {Barnwell}}, \ and\ \bibinfo {author}
  {\bibfnamefont {J.}~\bibnamefont {Zhang}},\ }\href {\doibase
  10.1063/1.3601927} {\bibfield  {journal} {\bibinfo  {journal} {Applied
  Physics Letters}\ }\textbf {\bibinfo {volume} {98}},\ \bibinfo {pages}
  {254101} (\bibinfo {year} {2011})}\BibitemShut {NoStop}%
\bibitem [{\citenamefont {Reinhold}\ \emph {et~al.}(2007)\citenamefont
  {Reinhold}, \citenamefont {Scholz}, \citenamefont {John},\ and\ \citenamefont
  {Hilleringmann}}]{Reinhold2007}%
  \BibitemOpen
  \bibfield  {author} {\bibinfo {author} {\bibfnamefont {C.}~\bibnamefont
  {Reinhold}}, \bibinfo {author} {\bibfnamefont {P.}~\bibnamefont {Scholz}},
  \bibinfo {author} {\bibfnamefont {W.}~\bibnamefont {John}}, \ and\ \bibinfo
  {author} {\bibfnamefont {U.}~\bibnamefont {Hilleringmann}},\ }\href@noop {}
  {\bibfield  {journal} {\bibinfo  {journal} {Journal of Communications}\
  }\textbf {\bibinfo {volume} {2}},\ \bibinfo {pages} {14} (\bibinfo {year}
  {2007})}\BibitemShut {NoStop}%
\bibitem [{\citenamefont {Yu}\ \emph {et~al.}(2011)\citenamefont {Yu},
  \citenamefont {Sandhu}, \citenamefont {Beiker}, \citenamefont {Sassoon},\
  and\ \citenamefont {Fan}}]{Yu2011}%
  \BibitemOpen
  \bibfield  {author} {\bibinfo {author} {\bibfnamefont {X.}~\bibnamefont
  {Yu}}, \bibinfo {author} {\bibfnamefont {S.}~\bibnamefont {Sandhu}}, \bibinfo
  {author} {\bibfnamefont {S.}~\bibnamefont {Beiker}}, \bibinfo {author}
  {\bibfnamefont {R.}~\bibnamefont {Sassoon}}, \ and\ \bibinfo {author}
  {\bibfnamefont {S.}~\bibnamefont {Fan}},\ }\href {\doibase 10.1063/1.3663576}
  {\bibfield  {journal} {\bibinfo  {journal} {Applied Physics Letters}\
  }\textbf {\bibinfo {volume} {99}},\ \bibinfo {pages} {214102} (\bibinfo
  {year} {2011})}\BibitemShut {NoStop}%
\bibitem [{\citenamefont {W.}, \citenamefont {III},\ and\ \citenamefont
  {L.}(1973)}]{W.1970}%
  \BibitemOpen
  \bibfield  {author} {\bibinfo {author} {\bibfnamefont {M. W.}\ \bibnamefont
  {Cardullo}},\bibinfo {author} {\bibfnamefont {W. L.}~\bibnamefont {Park III}},\ }\href@noop {}
  {\enquote {\bibinfo {title} {{US Patent, 3,713,148}},}\ }
  (\bibinfo {year} {1973})\BibitemShut {NoStop}%
\bibitem [{\citenamefont {Zink}\ and\ \citenamefont {Skuro}(2002)}]{Zink2002}%
  \BibitemOpen
  \bibfield  {author} {\bibinfo {author} {\bibfnamefont {U.}~\bibnamefont
  {Zink}}\ and\ \bibinfo {author} {\bibfnamefont {G.}~\bibnamefont {Skuro}},\
  }\href@noop {} {\enquote {\bibinfo {title} {{US Patent 6,661,197 B2}},}\ } (\bibinfo {year} {2003})\BibitemShut {NoStop}%
\bibitem [{\citenamefont {Bruning}(2008)}]{Bruning2008}%
  \BibitemOpen
  \bibfield  {author} {\bibinfo {author} {\bibfnamefont {G. W.}\ \bibnamefont
  {Bruning}},\ }\href@noop {} {\enquote {\bibinfo {title} {{US Patent 2003/0231001 A1}},}\ } (\bibinfo {year} {2008})\BibitemShut {NoStop}%
\bibitem [{\citenamefont {Ramrakhyani}, \citenamefont {Member},\ and\
  \citenamefont {Mirabbasi}(2011)}]{Ramrakhyani2011}%
  \BibitemOpen
  \bibfield  {author} {\bibinfo {author} {\bibfnamefont {A.~K.}\ \bibnamefont
  {Ramrakhyani}}, \bibinfo {author} {\bibfnamefont {S.}~\bibnamefont {Member}},
  \ and\ \bibinfo {author} {\bibfnamefont {S.}~\bibnamefont {Mirabbasi}},\
  }\href@noop {} {\bibfield  {journal} {\bibinfo  {journal} {IEEE Transaction
  on Biomedical Circuits and Systems}\ }\textbf {\bibinfo {volume} {5}},\
  \bibinfo {pages} {48} (\bibinfo {year} {2011})}\BibitemShut {NoStop}%
\bibitem [{\citenamefont {Waffenschmidt}\ and\ \citenamefont
  {Staring}(2009)}]{Waffenschmidt}%
  \BibitemOpen
  \bibfield  {author} {\bibinfo {author} {\bibfnamefont {E.}~\bibnamefont
  {Waffenschmidt}}\ and\ \bibinfo {author} {\bibfnamefont {T.}~\bibnamefont
  {Staring}},\ }\href@noop {} {\bibfield  {journal} {\bibinfo  {journal} {EPE
  '09. 13th European Conference on Power Electronics and Applications}\ ,\
  \bibinfo {pages} {1}} (\bibinfo {year} {2009})}\BibitemShut {NoStop}%
\bibitem [{\citenamefont {Kurs}\ \emph {et~al.}(2007)\citenamefont {Kurs},
  \citenamefont {Karalis}, \citenamefont {Moffatt}, \citenamefont
  {Joannopoulos}, \citenamefont {Fisher}, \citenamefont {Soljac},\ and\
  \citenamefont {Soljacic}}]{Kurs2007}%
  \BibitemOpen
  \bibfield  {author} {\bibinfo {author} {\bibfnamefont {A.}~\bibnamefont
  {Kurs}}, \bibinfo {author} {\bibfnamefont {A.}~\bibnamefont {Karalis}},
  \bibinfo {author} {\bibfnamefont {R.}~\bibnamefont {Moffatt}}, \bibinfo
  {author} {\bibfnamefont {J.~D.}\ \bibnamefont {Joannopoulos}}, \bibinfo
  {author} {\bibfnamefont {P.}~\bibnamefont {Fisher}}, \bibinfo {author}
  {\bibfnamefont {M.}~\bibnamefont {Soljac}}, \ and\ \bibinfo {author}
  {\bibfnamefont {M.}~\bibnamefont {Soljacic}},\ }\href {\doibase
  10.1126/science.1143254} {\bibfield  {journal} {\bibinfo  {journal}
  {Science}\ }\textbf {\bibinfo {volume} {317}},\ \bibinfo {pages} {83}
  (\bibinfo {year} {2007})}\BibitemShut {NoStop}%
\bibitem [{\citenamefont {Sample}, \citenamefont {Meyer},\ and\ \citenamefont
  {Smith}(2011)}]{Sample2011}%
  \BibitemOpen
  \bibfield  {author} {\bibinfo {author} {\bibfnamefont {A.~P.}\ \bibnamefont
  {Sample}}, \bibinfo {author} {\bibfnamefont {D.~A.}\ \bibnamefont {Meyer}}, \
  and\ \bibinfo {author} {\bibfnamefont {J.~R.}\ \bibnamefont {Smith}},\ }\href
  {\doibase 10.1109/TIE.2010.2046002} {\bibfield  {journal} {\bibinfo
  {journal} {IEEE Transactions on Industrial Electronics}\ }\textbf {\bibinfo
  {volume} {58}},\ \bibinfo {pages} {544} (\bibinfo {year} {2011})}\BibitemShut
  {NoStop}%
\bibitem [{\citenamefont {Si}\ \emph {et~al.}(2008)\citenamefont {Si},
  \citenamefont {Hu}, \citenamefont {Malpas},\ and\ \citenamefont
  {Budgett}}]{Si2008}%
  \BibitemOpen
  \bibfield  {author} {\bibinfo {author} {\bibfnamefont {P.}~\bibnamefont
  {Si}}, \bibinfo {author} {\bibfnamefont {A.~P.}\ \bibnamefont {Hu}}, \bibinfo
  {author} {\bibfnamefont {S.}~\bibnamefont {Malpas}}, \ and\ \bibinfo {author}
  {\bibfnamefont {D.}~\bibnamefont {Budgett}},\ }\href {\doibase
  10.1109/TBCAS.2008.918284} {\bibfield  {journal} {\bibinfo  {journal} {IEEE
  Transactions on Biomedical Circuits and Systems}\ }\textbf {\bibinfo {volume}
  {2}},\ \bibinfo {pages} {22} (\bibinfo {year} {2008})}\BibitemShut {NoStop}%
\bibitem [{\citenamefont {Hasegawa}(2004)}]{Hasegawa2004}%
  \BibitemOpen
  \bibfield  {author} {\bibinfo {author} {\bibfnamefont {R.}~\bibnamefont
  {Hasegawa}},\ }\href {\doibase 10.1016/j.msea.2003.10.258} {\bibfield
  {journal} {\bibinfo  {journal} {Materials Science and Engineering A}\
  }\textbf {\bibinfo {volume} {375-377}},\ \bibinfo {pages} {90} (\bibinfo
  {year} {2004})}\BibitemShut {NoStop}%
\bibitem [{\citenamefont {Wiltshire}\ \emph {et~al.}(2001)\citenamefont
  {Wiltshire}, \citenamefont {Pendry}, \citenamefont {Young}, \citenamefont
  {Larkman}, \citenamefont {Gilderdale},\ and\ \citenamefont
  {Hajnal}}]{Wiltshire2001}%
  \BibitemOpen
  \bibfield  {author} {\bibinfo {author} {\bibfnamefont {M.~C.}\ \bibnamefont
  {Wiltshire}}, \bibinfo {author} {\bibfnamefont {J.~B.}\ \bibnamefont
  {Pendry}}, \bibinfo {author} {\bibfnamefont {I.~R.}\ \bibnamefont {Young}},
  \bibinfo {author} {\bibfnamefont {D.~J.}\ \bibnamefont {Larkman}}, \bibinfo
  {author} {\bibfnamefont {D.~J.}\ \bibnamefont {Gilderdale}}, \ and\ \bibinfo
  {author} {\bibfnamefont {J.~V.}\ \bibnamefont {Hajnal}},\ }\href {\doibase
  10.1126/science.291.5505.849} {\bibfield  {journal} {\bibinfo  {journal}
  {Science}\ }\textbf {\bibinfo {volume} {291}},\ \bibinfo {pages} {849}
  (\bibinfo {year} {2001})}\BibitemShut {NoStop}%
\bibitem [{\citenamefont {Freire}, \citenamefont {Marques},\ and\ \citenamefont
  {Jelinek}(2008)}]{Freire2008c}%
  \BibitemOpen
  \bibfield  {author} {\bibinfo {author} {\bibfnamefont {M.~J.}\ \bibnamefont
  {Freire}}, \bibinfo {author} {\bibfnamefont {R.}~\bibnamefont {Marques}}, \
  and\ \bibinfo {author} {\bibfnamefont {L.}~\bibnamefont {Jelinek}},\ }\href
  {\doibase 10.1063/1.3043725} {\bibfield  {journal} {\bibinfo  {journal}
  {Applied Physics Letters}\ }\textbf {\bibinfo {volume} {93}},\ \bibinfo
  {pages} {231108} (\bibinfo {year} {2008})}\BibitemShut {NoStop}%
\bibitem [{\citenamefont {Shamonina}\ and\ \citenamefont
  {Solymar}(2004)}]{Shamonina2004}%
  \BibitemOpen
  \bibfield  {author} {\bibinfo {author} {\bibfnamefont {E.}~\bibnamefont
  {Shamonina}}\ and\ \bibinfo {author} {\bibfnamefont {L.}~\bibnamefont
  {Solymar}},\ }\href {\doibase 10.1088/0022-3727/37/3/008} {\bibfield
  {journal} {\bibinfo  {journal} {Journal of Physics D: Applied Physics}\
  }\textbf {\bibinfo {volume} {37}},\ \bibinfo {pages} {362} (\bibinfo {year}
  {2004})}\BibitemShut {NoStop}%
\bibitem [{\citenamefont {Veselago}(1968)}]{Veselago1968}%
  \BibitemOpen
  \bibfield  {author} {\bibinfo {author} {\bibfnamefont {V.}~\bibnamefont
  {Veselago}},\ }\href@noop {} {\bibfield  {journal} {\bibinfo  {journal}
  {Physics-Uspekhi}\ }\textbf {\bibinfo {volume} {10}},\ \bibinfo {pages} {509}
  (\bibinfo {year} {1968})}\BibitemShut {NoStop}%
\bibitem [{\citenamefont {Pendry}(2000)}]{Pendry2000}%
  \BibitemOpen
  \bibfield  {author} {\bibinfo {author} {\bibfnamefont {J.}~\bibnamefont
  {Pendry}},\ }\href {http://www.ncbi.nlm.nih.gov/pubmed/11041972} {\bibfield
  {journal} {\bibinfo  {journal} {Physical review letters}\ }\textbf {\bibinfo
  {volume} {85}},\ \bibinfo {pages} {3966} (\bibinfo {year}
  {2000})}\BibitemShut {NoStop}%
\bibitem [{\citenamefont {Cummer}(2003)}]{Cummer2003}%
  \BibitemOpen
  \bibfield  {author} {\bibinfo {author} {\bibfnamefont {S.~A.}\ \bibnamefont
  {Cummer}},\ }\href {\doibase 10.1063/1.1554778} {\bibfield  {journal}
  {\bibinfo  {journal} {Applied Physics Letters}\ }\textbf {\bibinfo {volume}
  {82}},\ \bibinfo {pages} {1503} (\bibinfo {year} {2003})}\BibitemShut
  {NoStop}%
\bibitem [{\citenamefont {Merlin}(2007)}]{Merlin2007}%
  \BibitemOpen
  \bibfield  {author} {\bibinfo {author} {\bibfnamefont {R.}~\bibnamefont
  {Merlin}},\ }\href {\doibase 10.1126/science.1143884} {\bibfield  {journal}
  {\bibinfo  {journal} {Science}\ }\textbf {\bibinfo {volume} {317}},\ \bibinfo
  {pages} {927} (\bibinfo {year} {2007})}\BibitemShut {NoStop}%
\bibitem [{\citenamefont {Grbic}, \citenamefont {Jiang},\ and\ \citenamefont
  {Merlin}(2008)}]{Grbic2008}%
  \BibitemOpen
  \bibfield  {author} {\bibinfo {author} {\bibfnamefont {A.}~\bibnamefont
  {Grbic}}, \bibinfo {author} {\bibfnamefont {L.}~\bibnamefont {Jiang}}, \ and\
  \bibinfo {author} {\bibfnamefont {R.}~\bibnamefont {Merlin}},\ }\href
  {\doibase 10.1126/science.1154753} {\bibfield  {journal} {\bibinfo  {journal}
  {Science}\ }\textbf {\bibinfo {volume} {320}},\ \bibinfo {pages} {511}
  (\bibinfo {year} {2008})}\BibitemShut {NoStop}%
\bibitem [{\citenamefont {Smith}\ \emph {et~al.}(2003)\citenamefont {Smith},
  \citenamefont {Schurig}, \citenamefont {Rosenbluth}, \citenamefont {Schultz},
  \citenamefont {Ramakrishna},\ and\ \citenamefont {Pendry}}]{Smith2003a}%
  \BibitemOpen
  \bibfield  {author} {\bibinfo {author} {\bibfnamefont {D.~R.}\ \bibnamefont
  {Smith}}, \bibinfo {author} {\bibfnamefont {D.}~\bibnamefont {Schurig}},
  \bibinfo {author} {\bibfnamefont {M.}~\bibnamefont {Rosenbluth}}, \bibinfo
  {author} {\bibfnamefont {S.}~\bibnamefont {Schultz}}, \bibinfo {author}
  {\bibfnamefont {S.~A.}\ \bibnamefont {Ramakrishna}}, \ and\ \bibinfo {author}
  {\bibfnamefont {J.~B.}\ \bibnamefont {Pendry}},\ }\href {\doibase
  10.1063/1.1554779} {\bibfield  {journal} {\bibinfo  {journal} {Applied
  Physics Letters}\ }\textbf {\bibinfo {volume} {82}},\ \bibinfo {pages} {1506}
  (\bibinfo {year} {2003})}\BibitemShut {NoStop}%
\bibitem [{\citenamefont {Fang}\ \emph {et~al.}(2005)\citenamefont {Fang},
  \citenamefont {Liu}, \citenamefont {Yen},\ and\ \citenamefont
  {Zhang}}]{Fang2005a}%
  \BibitemOpen
  \bibfield  {author} {\bibinfo {author} {\bibfnamefont {N.}~\bibnamefont
  {Fang}}, \bibinfo {author} {\bibfnamefont {Z.}~\bibnamefont {Liu}}, \bibinfo
  {author} {\bibfnamefont {T.}~\bibnamefont {Yen}}, \ and\ \bibinfo {author}
  {\bibfnamefont {X.}~\bibnamefont {Zhang}},\ }\href {\doibase
  10.1007/s00339-004-3160-6} {\bibfield  {journal} {\bibinfo  {journal}
  {Applied Physics A}\ }\textbf {\bibinfo {volume} {80}},\ \bibinfo {pages}
  {1315} (\bibinfo {year} {2005})}\BibitemShut {NoStop}%
\bibitem [{\citenamefont {Cui}\ \emph {et~al.}(2005)\citenamefont {Cui},
  \citenamefont {Cheng}, \citenamefont {Lu}, \citenamefont {Jiang},\ and\
  \citenamefont {Kong}}]{Cui2005}%
  \BibitemOpen
  \bibfield  {author} {\bibinfo {author} {\bibfnamefont {T.}~\bibnamefont
  {Cui}}, \bibinfo {author} {\bibfnamefont {Q.}~\bibnamefont {Cheng}}, \bibinfo
  {author} {\bibfnamefont {W.}~\bibnamefont {Lu}}, \bibinfo {author}
  {\bibfnamefont {Q.}~\bibnamefont {Jiang}}, \ and\ \bibinfo {author}
  {\bibfnamefont {J.}~\bibnamefont {Kong}},\ }\href {\doibase
  10.1103/PhysRevB.71.045114} {\bibfield  {journal} {\bibinfo  {journal}
  {Physical Review B}\ }\textbf {\bibinfo {volume} {71}},\ \bibinfo {pages} {1}
  (\bibinfo {year} {2005})}\BibitemShut {NoStop}%
\bibitem [{\citenamefont {Teo}, \citenamefont {Huang},\ and\ \citenamefont
  {Wang}(2011)}]{Teo2011}%
  \BibitemOpen
  \bibfield  {author} {\bibinfo {author} {\bibfnamefont {K. H.}\ \bibnamefont
  {Teo}}, \bibinfo {author} {\bibfnamefont {D.}~\bibnamefont {Huang}}, \ and\
  \bibinfo {author} {\bibfnamefont {B.}~\bibnamefont {Wang}},\ }\href@noop {}
  {\enquote {\bibinfo {title} {{US Patent 2011/0133564 A1}},}\ } (\bibinfo {year} {2011})\BibitemShut {NoStop}%
\bibitem [{\citenamefont {Urzhumov}\ and\ \citenamefont
  {Smith}(2011)}]{Urzhumov2011}%
  \BibitemOpen
  \bibfield  {author} {\bibinfo {author} {\bibfnamefont {Y.}~\bibnamefont
  {Urzhumov}}\ and\ \bibinfo {author} {\bibfnamefont {D.~R.}\ \bibnamefont
  {Smith}},\ }\href {\doibase 10.1103/PhysRevB.83.205114} {\bibfield  {journal}
  {\bibinfo  {journal} {Physical Review B}\ }\textbf {\bibinfo {volume} {83}},\
  \bibinfo {pages} {31} (\bibinfo {year} {2011})}\BibitemShut {NoStop}%
\bibitem [{\citenamefont {Ghali}\ and\ \citenamefont
  {Rahman}(2009)}]{Ghali2009}%
  \BibitemOpen
  \bibfield  {author} {\bibinfo {author} {\bibfnamefont {H.~A.}\ \bibnamefont
  {Ghali}}\ and\ \bibinfo {author} {\bibfnamefont {H.~A.}\ \bibnamefont
  {Rahman}},\ }in\ \href@noop {} {\emph {\bibinfo {booktitle} {Proceedings of
  the COMSOL conference 2009}}}\ (\bibinfo {address} {Milan},\ \bibinfo {year}
  {2009})\BibitemShut {NoStop}%
\bibitem [{\citenamefont {Grover}(1982)}]{Grover1982}%
  \BibitemOpen
  \bibfield  {author} {\bibinfo {author} {\bibfnamefont {F.~W.}\ \bibnamefont
  {Grover}},\ }\href
  {http://www.amazon.com/Inductance-Calculations-Working-Formulas-Tables/dp/08%
76645570} {\emph {\bibinfo {title} {{Inductance Calculations: Working Formulas
  and Tables}}}}\ (\bibinfo  {publisher} {Instrumentation Systems \&},\
  \bibinfo {year} {1982})\ p.\ \bibinfo {pages} {300}\BibitemShut {NoStop}%
\bibitem [{\citenamefont {Urzhumov}\ \emph {et~al.}(2011)\citenamefont
  {Urzhumov}, \citenamefont {Chen}, \citenamefont {Bingham}, \citenamefont
  {Padilla},\ and\ \citenamefont {Smith}}]{Urzhumov}%
  \BibitemOpen
  \bibfield  {author} {\bibinfo {author} {\bibfnamefont {Y.}~\bibnamefont
  {Urzhumov}}, \bibinfo {author} {\bibfnamefont {W.}~\bibnamefont {Chen}},
  \bibinfo {author} {\bibfnamefont {C.}~\bibnamefont {Bingham}}, \bibinfo
  {author} {\bibfnamefont {W.}~\bibnamefont {Padilla}}, \ and\ \bibinfo
  {author} {\bibfnamefont {D.~R.}\ \bibnamefont {Smith}},\ }  {\bibfield  {journal}
  {\bibinfo  {journal} {Phys. Rev. B}\ }\textbf {\bibinfo {volume}
  {85}},\ \bibinfo {pages} {054430} (\bibinfo {year} {2003})}\BibitemShut
  {NoStop}%
\bibitem [{\citenamefont {Shvets}(2003)}]{Shvets2003}%
  \BibitemOpen
  \bibfield  {author} {\bibinfo {author} {\bibfnamefont {G.}~\bibnamefont
  {Shvets}},\ }\href {\doibase 10.1117/12.509970} {\bibfield  {journal}
  {\bibinfo  {journal} {Proceedings of SPIE}\ }\textbf {\bibinfo {volume}
  {5221}},\ \bibinfo {pages} {124} (\bibinfo {year} {2003})}\BibitemShut
  {NoStop}%
\bibitem [{\citenamefont {Freire}\ \emph {et~al.}(2010)\citenamefont {Freire},
  \citenamefont {Jelinek}, \citenamefont {Marques},\ and\ \citenamefont
  {Lapine}}]{Freire2010}%
  \BibitemOpen
  \bibfield  {author} {\bibinfo {author} {\bibfnamefont {M.~J.}\ \bibnamefont
  {Freire}}, \bibinfo {author} {\bibfnamefont {L.}~\bibnamefont {Jelinek}},
  \bibinfo {author} {\bibfnamefont {R.}~\bibnamefont {Marques}}, \ and\
  \bibinfo {author} {\bibfnamefont {M.}~\bibnamefont {Lapine}},\ }\href
  {\doibase 10.1016/j.jmr.2009.12.005} {\bibfield  {journal} {\bibinfo
  {journal} {Journal of magnetic resonance (San Diego, Calif. : 1997)}\
  }\textbf {\bibinfo {volume} {203}},\ \bibinfo {pages} {81} (\bibinfo {year}
  {2010})}\BibitemShut {NoStop}%
\bibitem [{\citenamefont {Chen}\ \emph {et~al.}(2011)\citenamefont {Chen},
  \citenamefont {Bingham}, \citenamefont {Mak}, \citenamefont {Caira},\ and\
  \citenamefont {Padilla}}]{Chen2011}%
  \BibitemOpen
  \bibfield  {author} {\bibinfo {author} {\bibfnamefont {W.~C.}\ \bibnamefont
  {Chen}}, \bibinfo {author} {\bibfnamefont {C.~M.}\ \bibnamefont {Bingham}},
  \bibinfo {author} {\bibfnamefont {K.~M.}\ \bibnamefont {Mak}}, \bibinfo
  {author} {\bibfnamefont {N.~W.}\ \bibnamefont {Caira}}, \ and\ \bibinfo
  {author} {\bibfnamefont {W.~J.}\ \bibnamefont {Padilla}},\ }\href@noop {}
  {\enquote {\bibinfo {title} {{Extremely Sub-wavelength Planar Magnetic
  Metamaterials}},}\ } (\bibinfo {year} {2011}),\ \bibinfo {note} {submitted},\
  \Eprint {http://arxiv.org/abs/arXiv:1111.2024v1} {arXiv:arXiv:1111.2024v1}
  \BibitemShut {NoStop}%
\bibitem [{\citenamefont {Roberts}, \citenamefont {Kundtz},\ and\ \citenamefont
  {Smith}(2009)}]{Roberts2009}%
  \BibitemOpen
  \bibfield  {author} {\bibinfo {author} {\bibfnamefont {D.~A.}\ \bibnamefont
  {Roberts}}, \bibinfo {author} {\bibfnamefont {N.}~\bibnamefont {Kundtz}}, \
  and\ \bibinfo {author} {\bibfnamefont {D.~R.}\ \bibnamefont {Smith}},\ }\href
  {http://www.ncbi.nlm.nih.gov/pubmed/19770868} {\bibfield  {journal} {\bibinfo
   {journal} {Optics express}\ }\textbf {\bibinfo {volume} {17}},\ \bibinfo
  {pages} {16535} (\bibinfo {year} {2009})}\BibitemShut {NoStop}%
\bibitem [{\citenamefont {Schurig}\ and\ \citenamefont
  {Smith}(2005)}]{Schurig2005}%
  \BibitemOpen
  \bibfield  {author} {\bibinfo {author} {\bibfnamefont {D.}~\bibnamefont
  {Schurig}}\ and\ \bibinfo {author} {\bibfnamefont {D.~R.}\ \bibnamefont
  {Smith}},\ }\href {\doibase 10.1088/1367-2630/7/1/162} {\bibfield  {journal}
  {\bibinfo  {journal} {New Journal of Physics}\ }\textbf {\bibinfo {volume}
  {162}},\ \bibinfo {pages} {0} (\bibinfo {year} {2005})}\BibitemShut {NoStop}%
\bibitem [{\citenamefont {Lee}\ \emph {et~al.}(2007)\citenamefont {Lee},
  \citenamefont {Liu}, \citenamefont {Xiong}, \citenamefont {Sun},\ and\
  \citenamefont {Zhang}}]{Lee2007}%
  \BibitemOpen
  \bibfield  {author} {\bibinfo {author} {\bibfnamefont {H.}~\bibnamefont
  {Lee}}, \bibinfo {author} {\bibfnamefont {Z.}~\bibnamefont {Liu}}, \bibinfo
  {author} {\bibfnamefont {Y.}~\bibnamefont {Xiong}}, \bibinfo {author}
  {\bibfnamefont {C.}~\bibnamefont {Sun}}, \ and\ \bibinfo {author}
  {\bibfnamefont {X.}~\bibnamefont {Zhang}},\ }\href
  {http://www.ncbi.nlm.nih.gov/pubmed/19550875} {\bibfield  {journal} {\bibinfo
   {journal} {Optics express}\ }\textbf {\bibinfo {volume} {15}},\ \bibinfo
  {pages} {15886} (\bibinfo {year} {2007})}\BibitemShut {NoStop}%
\bibitem [{\citenamefont {Yao}\ \emph {et~al.}(2011)\citenamefont {Yao},
  \citenamefont {Yang}, \citenamefont {Yin}, \citenamefont {Bartal},\ and\
  \citenamefont {Zhang}}]{Yao2011}%
  \BibitemOpen
  \bibfield  {author} {\bibinfo {author} {\bibfnamefont {J.}~\bibnamefont
  {Yao}}, \bibinfo {author} {\bibfnamefont {X.}~\bibnamefont {Yang}}, \bibinfo
  {author} {\bibfnamefont {X.}~\bibnamefont {Yin}}, \bibinfo {author}
  {\bibfnamefont {G.}~\bibnamefont {Bartal}}, \ and\ \bibinfo {author}
  {\bibfnamefont {X.}~\bibnamefont {Zhang}},\ }\href {\doibase
  10.1073/pnas.1104418108} {\bibfield  {journal} {\bibinfo  {journal}
  {Proceedings of the National Academy of Sciences of the United States of
  America}\ }\textbf {\bibinfo {volume} {108}},\ \bibinfo {pages} {11327}
  (\bibinfo {year} {2011})}\BibitemShut {NoStop}%
\bibitem [{\citenamefont {Smith}\ and\ \citenamefont
  {Schurig}(2003)}]{Smith2003}%
  \BibitemOpen
  \bibfield  {author} {\bibinfo {author} {\bibfnamefont {D.~R.}\ \bibnamefont
  {Smith}}\ and\ \bibinfo {author} {\bibfnamefont {D.}~\bibnamefont
  {Schurig}},\ }\href {\doibase 10.1103/PhysRevLett.90.077405} {\bibfield
  {journal} {\bibinfo  {journal} {Physical Review Letters}\ }\textbf {\bibinfo
  {volume} {90}},\ \bibinfo {pages} {077405} (\bibinfo {year} {2003})}\BibitemShut
  {NoStop}%
\bibitem [{\citenamefont {Al\`{u}}\ and\ \citenamefont
  {Engheta}(2005)}]{Alu2005}%
  \BibitemOpen
  \bibfield  {author} {\bibinfo {author} {\bibfnamefont {A.}~\bibnamefont
  {Al\`{u}}}\ and\ \bibinfo {author} {\bibfnamefont {N.}~\bibnamefont
  {Engheta}},\ }\href@noop {} {\bibfield  {journal} {\bibinfo  {journal}
  {Journal of Applied Physics}\ }\textbf {\bibinfo {volume} {97}} (\bibinfo
  {year} {2005})}\BibitemShut {NoStop}%
\bibitem [{\citenamefont {Schurig}\ and\ \citenamefont
  {Smith}(2003)}]{Schurig2003}%
  \BibitemOpen
  \bibfield  {author} {\bibinfo {author} {\bibfnamefont {D.}~\bibnamefont
  {Schurig}}\ and\ \bibinfo {author} {\bibfnamefont {D.~R.}\ \bibnamefont
  {Smith}},\ }\href {\doibase 10.1063/1.1562344} {\bibfield  {journal}
  {\bibinfo  {journal} {Applied Physics Letters}\ }\textbf {\bibinfo {volume}
  {82}},\ \bibinfo {pages} {2215} (\bibinfo {year} {2003})}\BibitemShut
  {NoStop}%
\bibitem [{\citenamefont {Balmain}, \citenamefont {L\"{u}ttgen},\ and\
  \citenamefont {Kremer}(2002)}]{Balmain2002}%
  \BibitemOpen
  \bibfield  {author} {\bibinfo {author} {\bibfnamefont {K.~G.}\ \bibnamefont
  {Balmain}}, \bibinfo {author} {\bibfnamefont {A.~A.~E.}\ \bibnamefont
  {L\"{u}ttgen}}, \ and\ \bibinfo {author} {\bibfnamefont {P.~C.}\ \bibnamefont
  {Kremer}},\ }\href@noop {} {\bibfield  {journal} {\bibinfo  {journal} {IEEE
  Antenna and Wireless Propagation Letters}\ }\textbf {\bibinfo {volume} {1}},\
  \bibinfo {pages} {146} (\bibinfo {year} {2002})}\BibitemShut {NoStop}%
\bibitem [{\citenamefont {Jacob}, \citenamefont {Alekseyev},\ and\
  \citenamefont {Narimanov}(2006)}]{Jacob2006}%
  \BibitemOpen
  \bibfield  {author} {\bibinfo {author} {\bibfnamefont {Z.}~\bibnamefont
  {Jacob}}, \bibinfo {author} {\bibfnamefont {L.~V.}\ \bibnamefont
  {Alekseyev}}, \ and\ \bibinfo {author} {\bibfnamefont {E.}~\bibnamefont
  {Narimanov}},\ }\href {\doibase 10.1364/OE.14.008247} {\bibfield  {journal}
  {\bibinfo  {journal} {Optics Express}\ }\textbf {\bibinfo {volume} {14}},\
  \bibinfo {pages} {8247} (\bibinfo {year} {2006})}\BibitemShut {NoStop}%
\bibitem [{\citenamefont {Smith}, \citenamefont {Kolinko},\ and\ \citenamefont
  {Schurig}(2004)}]{Smith2004a}%
  \BibitemOpen
  \bibfield  {author} {\bibinfo {author} {\bibfnamefont {D.~R.}\ \bibnamefont
  {Smith}}, \bibinfo {author} {\bibfnamefont {P.}~\bibnamefont {Kolinko}}, \
  and\ \bibinfo {author} {\bibfnamefont {D.}~\bibnamefont {Schurig}},\ }\href
  {\doibase 10.1364/JOSAB.21.001032} {\bibfield  {journal} {\bibinfo  {journal}
  {Journal of the Optical Society of America B}\ }\textbf {\bibinfo {volume}
  {21}},\ \bibinfo {pages} {1032} (\bibinfo {year} {2004})}\BibitemShut
  {NoStop}%
\bibitem [{\citenamefont {Smith}\ \emph {et~al.}(2004)\citenamefont {Smith},
  \citenamefont {Schurig}, \citenamefont {Mock}, \citenamefont {Kolinko},\ and\
  \citenamefont {Rye}}]{Smith2004b}%
  \BibitemOpen
  \bibfield  {author} {\bibinfo {author} {\bibfnamefont {D.~R.}\ \bibnamefont
  {Smith}}, \bibinfo {author} {\bibfnamefont {D.}~\bibnamefont {Schurig}},
  \bibinfo {author} {\bibfnamefont {J.~J.}\ \bibnamefont {Mock}}, \bibinfo
  {author} {\bibfnamefont {P.}~\bibnamefont {Kolinko}}, \ and\ \bibinfo
  {author} {\bibfnamefont {P.}~\bibnamefont {Rye}},\ }\href {\doibase
  10.1063/1.1690471} {\bibfield  {journal} {\bibinfo  {journal} {Applied
  Physics Letters}\ }\textbf {\bibinfo {volume} {84}},\ \bibinfo {pages} {2244}
  (\bibinfo {year} {2004})}\BibitemShut {NoStop}%
\bibitem [{\citenamefont {Siddiqui}\ and\ \citenamefont
  {Eleftheriades}(2011)}]{Siddiqui2011}%
  \BibitemOpen
  \bibfield  {author} {\bibinfo {author} {\bibfnamefont {O.~F.}\ \bibnamefont
  {Siddiqui}}\ and\ \bibinfo {author} {\bibfnamefont {G.~V.}\ \bibnamefont
  {Eleftheriades}},\ }\href {\doibase 10.1016/j.jfranklin.2010.02.005}
  {\bibfield  {journal} {\bibinfo  {journal} {Journal of the Franklin
  Institute}\ }\textbf {\bibinfo {volume} {348}},\ \bibinfo {pages} {1285}
  (\bibinfo {year} {2011})}\BibitemShut {NoStop}%
\bibitem [{\citenamefont {Cheng}\ \emph {et~al.}(2008)\citenamefont {Cheng},
  \citenamefont {Liu}, \citenamefont {Mock}, \citenamefont {Cui},\ and\
  \citenamefont {Smith}}]{Cheng2008a}%
  \BibitemOpen
  \bibfield  {author} {\bibinfo {author} {\bibfnamefont {Q.}~\bibnamefont
  {Cheng}}, \bibinfo {author} {\bibfnamefont {R.}~\bibnamefont {Liu}}, \bibinfo
  {author} {\bibfnamefont {J.}~\bibnamefont {Mock}}, \bibinfo {author}
  {\bibfnamefont {T.}~\bibnamefont {Cui}}, \ and\ \bibinfo {author}
  {\bibfnamefont {D.~R.}\ \bibnamefont {Smith}},\ }\href {\doibase
  10.1103/PhysRevB.78.121102} {\bibfield  {journal} {\bibinfo  {journal}
  {Physical Review B}\ }\textbf {\bibinfo {volume} {78}},\ \bibinfo {pages} {121102}
  (\bibinfo {year} {2008})}\BibitemShut {NoStop}%
  
\end{thebibliography}
\end{document}